\documentclass[5p]{elsarticle}
\usepackage{amssymb}
\usepackage{amsmath}
\usepackage{graphicx}
\usepackage[table]{xcolor}
\usepackage{epsfig}
\usepackage[utf8]{inputenc}
\usepackage{dcolumn}
\usepackage{bm}
\usepackage{array}
\usepackage{multirow} 
\usepackage{lineno,hyperref}
\modulolinenumbers[5]
\biboptions{sort&compress}

\begin{document}

\title{Tuning the Transdermal Transport by Application of External Continuous Electric Field: A Molecular Dynamics Coarse-Grained Study}

\author{Neila Machado$^{1}$, Clarissa Callegaro$^{2}$, Marcelo Augusto Christoffolete$^{1}$,  and Herculano Martinho\corref{mycorrespondingauthor}}

\address{$^{1}$Centro de Ciências Naturais e Humanas, Universidade Federal do ABC, Av. dos Estados 5001, Santo André-SP, 09210-580, Brazil\\
$^{2}$AXCEM Dermatologia, Cosmiatria e Laser, R. Taváres Cabral, 102, conj. 71, São Paulo-SP, 05423-030, Brazil }

\cortext[mycorrespondingauthor]{Corresponding author: herculano.martinho@ufabc.edu.br}

\begin{abstract}

Since a long time the application of small electric potentials on biological membranes (iontophoresis) proved enabling control and improvement of transdermal delivery of substances across this barrier. In spite of a large experimental data, the detailed molecular mechanism of iontophoresis is absent. In the present work the interaction among the external continuous electric field with the outermost layer of the skin (\textit{stratum corneum}) was studied by coarse-grained molecular dynamics. Our results pointed out the occurrence of water-rich vesicles formation depending on the field strength. The corresponding phase diagram indicated that the large set of phenomena (vesicle formation, reversibility, phase transition, disruption) could be completely controlled by tuning external continuous electric fields. Interestingly,  electric field shielding effects are in the origin of observed effects and followed a general Arrhenius-like time dependence. Direct current (DC) electric field usage would also have booster diffusion effects due to vesicles creation and reincorporation which would have direct beneficial absorption effects on water-soluble topical agents (vitamins) or dermal jet-injection of drugs by fine needles.


\end{abstract}

\maketitle

\section{Introduction}

The human skin is a natural barrier to the transport of substances. Skin controls the detailed balance of, e.g.,  moisture and harmful molecules. It protects the body against external chemical and physical factors, takes part in the metabolic processes, plays a resorptive and thermoregulatory function, and it partakes in immunological processes\cite{hongbo,boer2016structural}. In fact, it is a complex organ that has many functions that go far beyond its role as a barrier to the environment\cite{hongbo}.

The basic structure of the skin comprises three component layers: epidermis, dermis, and hypodermis\cite{Burkitt2014}. The epidermis is the outer layer of the three layers that make up the skin\cite{Burkitt2014}. It is divided into four layers, starting at the dermal junction with the basal cell layer and ending at the outer surface in the \textit{stratum corneum} (SC) membrane\cite{JamesG.MarksJr.2013}. SC is the main responsible for the selective permeability of exogenous substances. The SC is organized in an arrangement known as "bricks and mortar" where the keratinized cells called corneocytes represent the "bricks" and are embedded into a highly specialized intercellular cement known as the extracellular matrix or lipid matrix\cite{Harding2013}. This structured bilayered matrix is mainly composed by ceramide, cholesterol, and free fatty acids molecules. The detailed composition and amount of bilayers vary as function of the depth of the SC\cite{hongbo, Burkitt2014, elias2016advances,JamesG.MarksJr.2013}. This multilamellar system is responsible for the maintenance and homeostatic balance of the skin\cite{hongbo, Burkitt2014, elias2016advances,JamesG.MarksJr.2013}.

The SC is a rate-limiting lipophilic barrier against the uptake of chemical and biological toxins, as well as trans-epidermal water loss\cite{fox2011transdermal,herman2015essential}. Several approaches have been proposed to control the permeability of this barrier (see, e.g., \cite{dkabrowska2018relationship} and \cite{pecoraro2019predicting}). Such control would enable to establish viable and advantageous optional routes for the administration of medications, vitamins, and  nutrients\cite{Wosicka2017}. The transdermal delivery could for example, reduce first-pass metabolism associated with oral delivery, and is not painful as injections\cite{schoellhammer2014skin}.

A number of factors affects the dermal absorption, including, e.g., type of skin; physical-chemical properties of delivery systems; skin moisture level; external temperature; skin pre-treatment; among others. Thus, numerous \textit{in-vitro} and \textit{in-vivo} models are used to examine the penetration of active compounds through the skin\cite{herman2015essential,godin2007transdermal}. Notwithstanding, several issues such as permeability and dermal absorption processes, remain unsolved.

The application of external electric fields in the biological membranes aiming enable the transport of substances across this barrier is known as \textit{electroporation}. The effect can be either reversible or irreversible\cite{BatistaNapotnik2018, kotnik2015electroporation}. The reversible electroporation has become an important tool in biotechnology and medicine. There are several reports considering the fusion of cells, treatment of cancer, and transdermal delivery of drugs and genes using reversible electroporation (see, e.g., \cite{kotnik2015electroporation}). Irreversible electroporation also have is also an important medical technology as, e.g., tissue ablation tool for treatment of elusive cancers such as in the prostate, the lung and the brain\cite{rubinsky2007irreversible}.

Furthermore, electroporation is widely used as a delivery method for small DNA transfection in cell culture as well\textit{in vivo}\cite{potter2018transfection}. It is also used to transform bacteria as standard procedure in molecular biology. It is reported to present superior efficiency compared to chemical or thermal shock techniques\cite{potter2018transfection}. Therefore, efficiency improvement in electroporation protocols is very desirable.

The electric field pulses usually employed experimentally have intensities of $\sim 1$ mV/nm and exposure duration of few milliseconds\cite{kotnik2015electroporation}. The temporary pores formed are difficult to be directly observed since their dimensions (several nanometers) and instability fall out the limits of current electron microscopy soft matter experimental techniques\cite{kotnik2015electroporation}. Thus, the concretely established characteristic of the electroporation include the creation of a transient charge state where there is a rapid and reversible increase in permeability, electrical conductance and a decrease in membrane resistance\cite{kotnik2015electroporation,BatistaNapotnik2018}.

\textit{In-silico} approaches are in principle able to predict adsorption, distribution, metabolism, excretion, and toxicity on the basis of input data describing physical-chemical properties of the compounds to be delivered and on physiological properties of the exposed skin\cite{Baba2015,Selzer2015,Hatanaka2015,Verheyen2017,marrink2019computational,Machado2016,rocco2017molecular,gupta2018electroporation}. In particular, classical molecular dynamics have been used to perform permeability studies at molecular level\cite{Machado2016,rocco2017molecular,gupta2018electroporation,marrink2019computational}. 

The change caused by pulsed electric fields on the lateral organization of cell membranes is widely reported on literature (see, e.g.,\cite{gupta2018electroporation,Groves2000,tarek2017atomistic}). The first work reporting a possible molecular mechanism for pore formation in the skin lipid bilayer during electroporation was the molecular dynamics simulation by Gupta and Rai\cite{gupta2018electroporation}. They studied the effect of the applied external electrical field ($0.6-1.0$ V/nm) on the pore formation dynamics in the lipid bilayer of different sizes (154, 616, and 2464 lipids) and compositions (ceramide/cholesterol/free fatty acid, 1:0:0, 1:0:1, 1:1:0, 1:1:1). They reported that the electroporation process was found to be reversible for the setted parameters. The authors also reported that the pore-opening dynamics depends upon the strength of the applied electric field, whereas the pore-closing dynamics is independent of the applied electric field. The interfacial water played a key role in the electroporation of the skin lipid bilayer. 

The external disturbances can lead to membrane re-organization at different levels, e.g., affecting organizational tendencies such as compartmentalization and groupings\cite{Groves2000}. In membranes with different amounts of cholesterol molecules in their composition, the application of electric fields leads to the formation of pores with different morphologies, when compared to the hydrophilic pores often formed in phospholipid membranes\cite{Polak2014,Koronkiewicz2004,Casciola2014}. These morphological differences have impact on the transport properties of such electroporated membranes. 

The simulation studies by Casciola \textit{et al.}\cite{Casciola2014} concluded that electroporation of cholesterol-rich membranes is strongly influenced by this compound in agreement with previous studies of Koronkiewicz and Kalinowski \cite{Koronkiewicz2004}. The former authors shown that the pores formed in membranes containing cholesterol close during electroporation and non-stabilized pores of the phospholipid head groups collapse and seal much more rapidly.

In the molecular dynamics studies investigating the thermodynamic effect on dipalmitoylphosphatidylcholine (DPPC) bilayers with different cholesterol concentrations, Bennet\textit{et al.}\cite{Bennett2009} reported that cholesterol was able to increase the order, thickness and stiffness of these bilayers, restricting deformations of the bilayer and preventing the formation of pores. They suggested that large thermal fluctuations are involved in the budding and fusion of vesicles, in the passive lipid flip-flop and pore formation. The increased concentration of this molecule promoted the increase of the free energy barrier to transfer the sets of DPPC heads to the center of the bilayer by lowering the velocity of the DPPC flip-flop in orders of magnitude.

Fernández \textit{et al.}\cite{Fernandez2010} calculated the electroporation field threshold on DOPC bilayers with different cholesterol concentrations. They reported that membranes containing cholesterol presented greater electric field threshold for electroporation, slower kinetics, and higher cohesion as well.

The application of continuous small electric potentials enables control and improvement of transdermal molecules delivery\cite{singhal2017iontophoresis,prausnitz2008transdermal}. This methodology is named iontophoresis and has medical applications for treatment and diagnosis in a broad range of pathologies as hyperhidrosis\cite{gollins2019retrospective}, ophthalmology\cite{prausnitz2008transdermal}, cystic fibrosis\cite{wallis2019diagnosis}, anaesthesia\cite{liu2017iontophoretic}, cancer \cite{byrne2018use}, among others (see, e.g.,\cite{gratieri2017iontophoresis}). The typical electrical currents can vary between $0.5$ and $30$ mA/cm$^{2}$ which is applied for minutes or hours\cite{kassan1996physical,banga1999iontophoresis,jadoul1999effects}. The technique enables transdermal administration and provides programmable drugs dosage by setting the electric current. It helps to make drug absorption less dependent on biological variables\cite{banga1999iontophoresis}
In spite of their relevance, to the best of our knowledge there are a lack of computational simulations studies concerning the molecular mechanisms of iontophoresis and electroporation as well. Its exact action mechanisms still obscure\cite{stewart2018intracellular}. Moreover, in order to have a complete picture about the action of external electric fields over biological membranes, the low intensity, continuous field is an important limit case to be studied. In this work the interaction of low intensity static electric fields and SC model of human skin was studied using molecular dynamics (MD).

\section{Methodology}

In this study,  MD was used in the coarse-grained (CG) approach\cite{Marrink2004,Marrink2007,Ingolfsson2014a}. This methodology allows the simulation of large and complex systems such as biological ones, providing a complete description that is beyond the dimensions usually explored in the actual soft matter experimental studies. All simulations were implemented in the GROMACS MD v.2018.1\cite{VanDerSpoel2005,Abraham2015,Berendsen1995}. 

\begin{figure}[th!]
\centering
	\includegraphics[width=7cm]{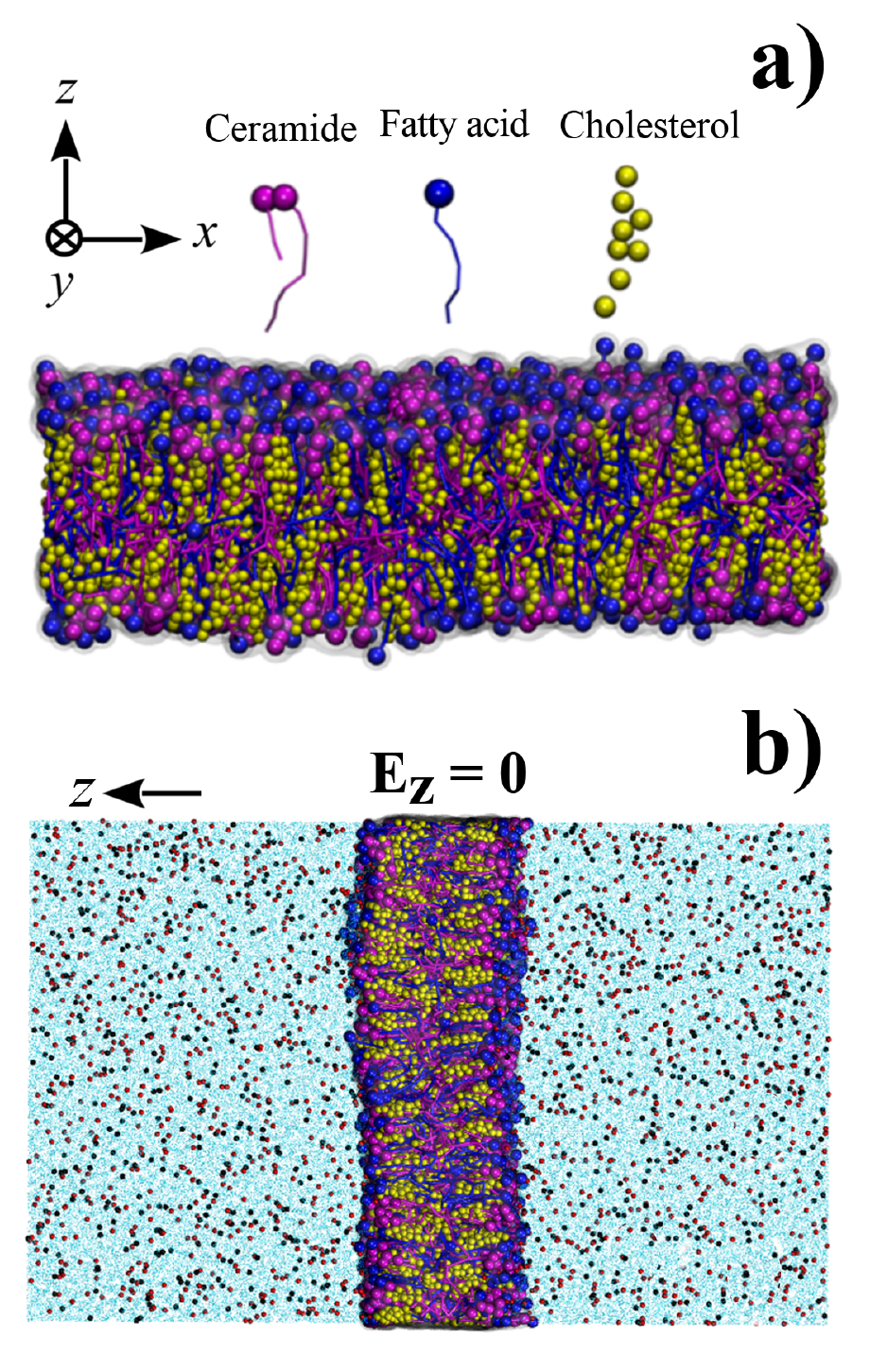}
	\caption{\textbf{a)} Coarse-grained representation for the human \textit{stratum corneum} lamellar lipidic membrane containing  ceramide ($24\%$), fatty acid ($39\%$), and cholesterol($36\%$). The membrane thickness is 5.4 nm. The depth of the bilayer was oriented along z-axis following the oriented axis set shown in the inset.  \textbf{b) }Initial state of the simulation box in the zero-field condition. The dimensions of box are 25 x 25 x 18 nm$^3$. The membrane is surrounded by water (cyan) and ions (Na$^{+}$, red dots and Cl$^{-}$, black dots) }\label{skinmodel}
\end{figure}

The SC model was  developed and describe elsewhere by one of the authors\cite{Machado2016}. It uses the composition of the lipid matrix region (lamellar region between the corneocytes): i) ceramide type II ($24\%$), which forms a dense bilayer and acts on the input control of water molecules making the membrane less permeable compared to the membrane of phospholipids; ii) cholesterol($36\%$), which acts as a molecular "clamp" between the larger molecules and which is also involved in membrane fluidity; iii) fatty acid (tetrasanoic acid, $39\%$) which decreases the packaging of the membrane (Fig. \ref{skinmodel}a). The polarized water is explicitly included in this SC model\cite{Yesylevskyy2010}. The membrane was modeled using the INSANE\cite{Wassenaar2015} program and contained $65,166$ polarized water beads. Na$^{+}$ and Cl$^{-}$ ions were incorporated into the system. Thus, the initial system totaled 69,462 molecules. The total volume of the simulation was defined in a box of $25\times 25 \times 18$ nm$^{3}$.

We used the Martini force-field\cite{Marrink2004,Marrink2007}. For integration of Newton's equations of motion it was used the leap-frog algorithm\cite{Hockney1974} in order to obtain the MD trajectories. The integration time step was 20 fs. The search for the neighborhood was using the Verlet scheme\cite{Verlet1967}. For the treatment of Coulomb forces we use the reaction-field/potential-shift with a cutting radius of $1.1$ nm. For the van der Waals interactions we used the cut-off/potential-shift with Lennard-Jones cutoff radius of $1.1$ nm. The dielectric constant was set to $15$ which is a value very close to the dielectric constant of membrane lipids \cite{Dilger1196, Gramse2013}.For temperature control we used the V-rescale thermostat (a modified Berendsen type)\cite{Berendsen1984,Bussi2007}. A temperature bath of 310 K was chosen due to its approximate value to the human physiological temperature. In order to control the pressure, the Parrinello-Rahman\cite{parrinello1981polymorphic} barostat was used to couple the system to semi-isotropic pressure of 1 atm with comprehensibility of $3\times 10^{-4}$  bar along the x- and y-axis directions. The axis between the bilayers was chosen along x-direction (see Fig.\ref{skinmodel}a). Periodic boundary conditions were applied in the directions of the x, y, and z axes.

The constant electric field was applied along the z-axis ($E_z$) using the GROMACS implementation by Caleman and Spoel\cite{caleman2008picosecond}. We notice that the induced electric field from the dipoles on the simulation box ($E_{dp}$) was always  $E_{dp}\sim 10^{-10}E_z$. Thus, the effects of $E_{dp}$ could be neglected. The temporal evolution in each situation was monitored, looking for frequent structural alterations. Before application of the electric field, the system was relaxed  by running 1000 ns of trajectory simulation. The minimized state resulted in a set box at 18.63 x 18.63 x 29.76 nm$^3$. It was used as the basis for the subsequent simulations and variations of electric field intensities. The results where visualized using the VMD program\cite{humphrey1996vmd}.

\section{Results and discussion}

We analyzed the effects of electric field in the range $E_{z}=0-100$ mV/nm. The results are presented and discussed on the following. A summary of findings is presented on Table \ref{table}.

\begin{table}[tbh!]
\small
\centering
\caption{Summary of findings in the vesicle dynamics as function of electric field strength.}\label{table}
\begin{tabular}{|p{1.5cm}|p{1.5cm}|p{4.5cm}|}
\hline 
$E_{z}$(mV/nm) & Total time (ns) & Qualitative Description \\ 
\hline 
\hline 
$ \leq6$ & $1000$ & absence of effects \\ 
\hline 
$7;8$ & $1000$ & waviness deformation in the membrane; water incorporation in the core; reversible iontophoresis\\ 
\hline 
$9$ & $1000$ & ripping off vesicle with water in the core; irreversible iontophoresis \\ 
\hline
$10$ & $1000$ & ripping off vesicle with water in the core; membrane unstable after vesicle re-incorporation; irreversible iontophoresis \\ 
\hline 
$20-100$ & $1-100$ & fragmentation of the membrane; irreversible iontophoresis \\ 
\hline
\end{tabular} 
\end{table}

\begin{figure*}[th!]
\centering
	\includegraphics[width=15cm]{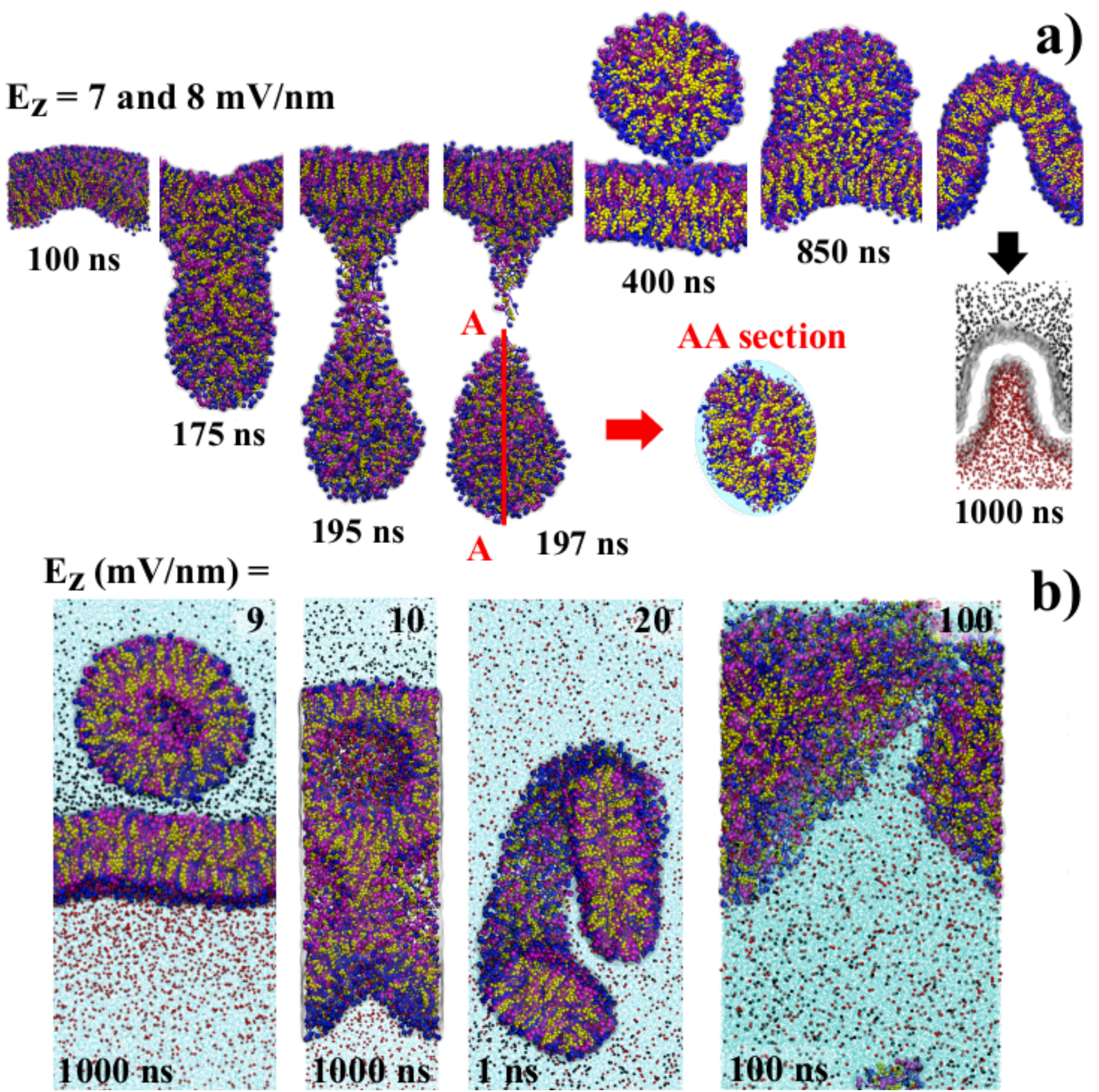}
	\caption{\textbf{a)} Temporal evolution up to $1000$ ns for z-axis applied electric fields  $E_{z} = 7$ and $8$ mV/nm. For these intensities the detachment/attachment of a vesicle was observed. The vesicle  encapsulates water molecules in the core and also in an internal shell close to the surface as shown in the $AA$ section of $197$ ns snapshot. The detachment was completed at this time. At $400$ ns the vesicle starts to be reincorporated. At $1000$ ns it was completely reabsorbed on the membrane. However, a charge separation was markedly observed as indicated by  the bottom frame for $1000$ ns. For these field intensities it was observed a reversible iontophoresis. \textbf{b)} Final simulation snapshots for $E_{z} = 9, 10, 20$ and $100$ mV/nm. For $9$ mV/mn electric field intensity it was observed the detachment of a vesicle which will not return to the membrane.  For  $E_{z} = 10$ mV/nm, a vesicle is expelled and returns to the membrane. However the final topology is unstable having a honeycomb pattern (phase transition). The process is irreversible. For $20$ mV/nm and $100$ mV/nm field intensities the membrane is deformed in different fragments also in a irreversible way. }\label{snapshots}
\end{figure*}

\subsection{Field strength $E_{z}\leq6$ mV/nm}

Our findings indicate absence of noticeable topological effects on the membrane in this case. Electric fields of this range of intensity only promoted the separation of ionic charges. Zero-field simulations (Fig. \ref{skinmodel}b) showed a homogeneous ions distribution in the solvent. It was observed that the membrane became more compact (reduction of $20$ \% in the box volume) as widely reported for membranes with cholesterol in the structure\cite{Koronkiewicz2004}.

\begin{figure*}[tbh!]
\centering
	\includegraphics[width=12cm]{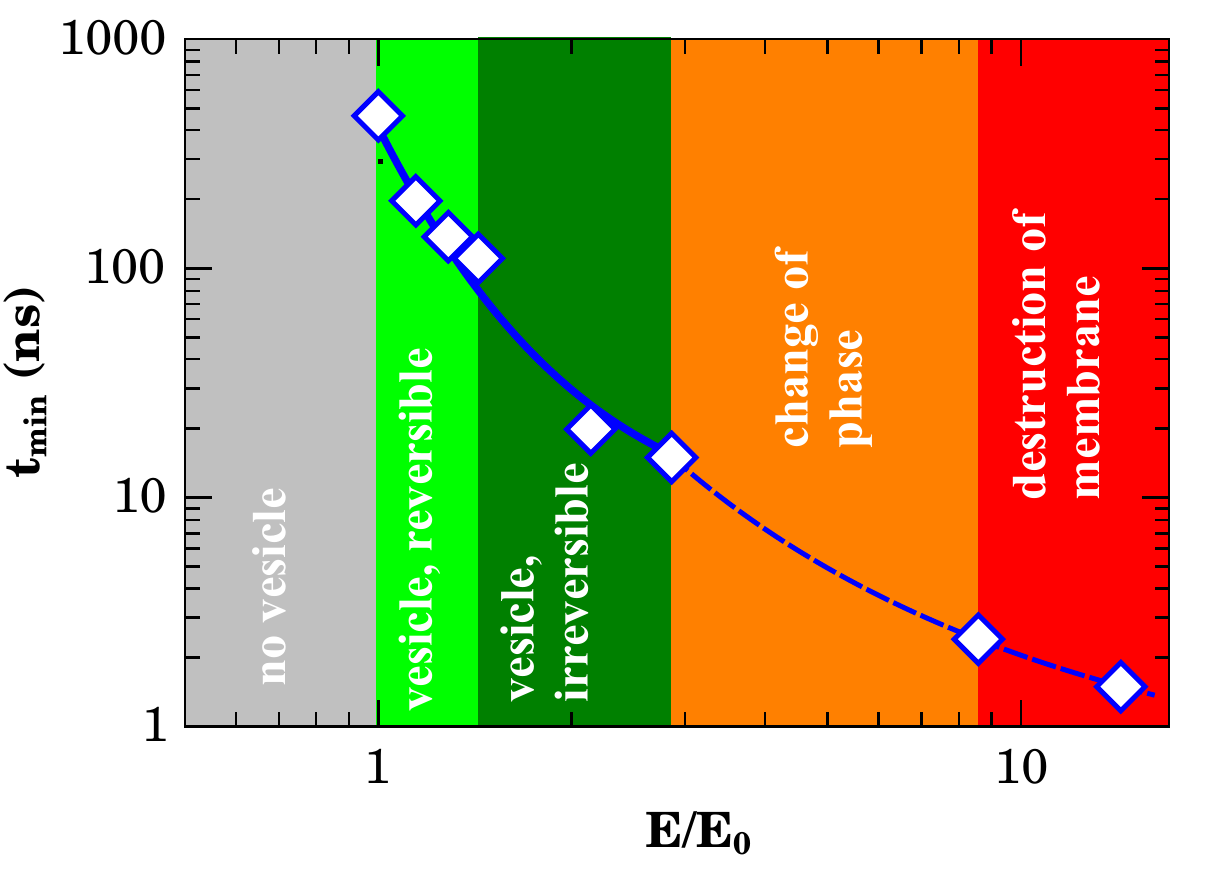}
	\caption{Diagram showing the minimal time for events for \textit{stratum corneum} membrane model ($t_{min}$) as function of the normalized static electric field strength ($E/E_{0}$). The normalizing factor is the minimal electric field where vesicles formation takes place in the simulation ($E_{0} = 7$ mV/nm in the present case). The solid and dashed lines are fittings to eq. (1) as described in the text.}\label{diagram}
\end{figure*}

Structural variations were also absent. In fact, only a slight bending promoted by the proximity of the ions that were rearranged where observed. Na$^{+}$ cations concentrated on the top of membrane (z-axis), and the Cl$^{-}$ anions on the bottom. The simulation extended beyond $1$ $\mu$s without noticeable changes (see supplementary material).

\subsection{Field strength $E_{z} = 7$ and $8$ mV/nm}

Applying  intensities of $7$ and $8$ mV/nm field we obtained very outstanding results. The starting deformation in the membrane (see $100$ ns snapshot in Fig. \ref{snapshots}a) resulted in the later formation of vesicle. This vesicle encapsulated water molecules in the core and in an internal shell close to the surface (see AA section of $197$ ns snapshot). The detachment was completed at this time. At $400$ ns the vesicle starts to be re-incorporated. At $1000$ ns it was completely reabsorbed on the membrane. However, a charge separation was markedly observed as indicated by the bottom frame for $1000$ ns. For these field intensities it was observed a reversible iontophoresis process.

It was also observed that the vesicle detachment/return occurred $\sim50$\% faster for $E_{z}=8$ mV/nm. Extending the simulations in both intensities up to $2000$ ns, we observed that the process of formation, detachment and re-absorption of the vesicles.

\subsection{Field strength $9 \leq E_{z} \leq 100 $ mV/nm}

The $E_z = 9$ mV/nm (Fig. \ref{snapshots}b) presented similar results of $7$ and $8$ mV/nm. Notwithstanding the resemblance, the formed vesicle did not return to the membrane. For $E_z = 10$ mV/nm (Fig. \ref{snapshots}b) vesicle enclosing water molecules was also formed but the membrane was unable to recover. The vesicle returned to the membrane, but the membrane did not stabilize. The final topology was unstable having a honeycomb pattern. Increasing field strength to $20 – 100$ mV/nm (Fig \ref{snapshots}b) resulted fragmentation of the SC membrane leading drastic and irreversible morphological changes. More figures and details considering another fields are available in the supplementary materials section.

The simulations reported on literature are usually based on membranes formed essentially of few kinds of phospholipids subjected to pulsed electric fields of various magnitudes. In general, the reported result is the formation of hydrophilic pores which are stabilized by the head groups of the phospholipids\cite{BatistaNapotnik2018,kotnik2015electroporation,rubinsky2007irreversible,Polak2014,Koronkiewicz2004,Casciola2014,Bennett2009}.

Our results shows the possibility of controlling both water transport across SC and vesicle formation dynamics without large scale membrane compromising by tuning the electric field intensity.  In order to summarize the data and give a global phenomenological explanation to the overall picture of the previous results we present the diagram of Fig. \ref{diagram}. In this diagram, the minimal time for relevant events (reversible or irreversible vesicle formation, change of phase or membrane disruption) was plotted against the applied field normalized to the minimal time for events starting ($E_0 = 7$ mV/nm in the present case) in a log-reciprocal scale. The normalization helps to future comparison to experimental data. The data was fitted to an Arrhenius-like dependence

\begin{equation}
t_{min}\varpropto e^{\frac{\alpha E}{E_0} }
\label{tmin}
\end{equation}

Due to the both internal and external charge separation to the vesicles and dielectric constant mismatch as well, the external field is shielded and the local field is decreased in relation to external one. One rough approximation to the effective local field is $E_{\alpha}=E_{0}/\alpha$ being $\alpha$ the shielding factor. The solid line shown on Fig.\ref{diagram} presents the best fitting to eq. \ref{tmin} for data in the vesicle formation region of the diagram ($1 \leq E/E_{0} \leq 1.9$). The accordance was excellent for $\alpha=5.7\pm 0.3$. In this case, $E_{\alpha}=0.175\times E_{0}=1.2$ mW/nm. For increasing fields strengths where change of phase or destruction of membrane takes place ($E/E_{0}\gtrsim 1.9$) the slope of the curve presented a slight change. The best fit (dashed line on Fig. \ref{diagram}) was obtained for $\alpha=6.7 \pm 0.5$ which furnished  $E_{\alpha}=0.149\times E_{0}=1.0$ mW/nm. For comparison effects, the electric field shielding due to the presence of a dielectric sphere is $\alpha\sim 5.3$\cite{heald1995classical} which is very similar to our results. Despite the broad values of intensities of applied fields a large diversity of membrane events in the diagram appeared to emerge from the same local field $E_{\alpha}\sim 1.1$ mV/nm. At first glance, one possible explanation could reside on a viscosity field dependence which would be critical at micro-scale where non-Newtonian phenomena arises\cite{pipe2009microfluidic}.

\begin{figure}[th!]
\centering
	\includegraphics[width=8cm]{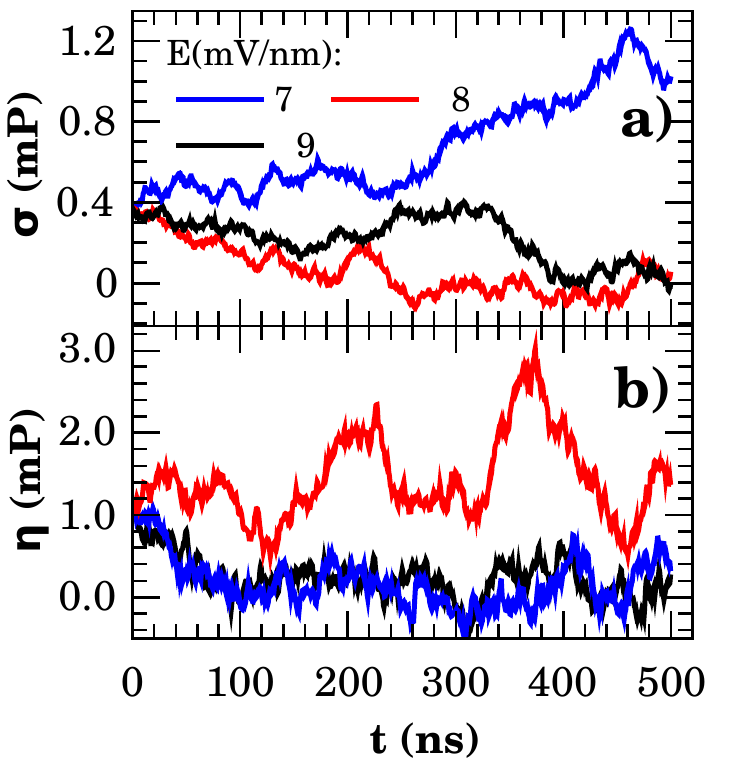}
	\caption{Shear (a) and bulk (b) viscosity of the system under electric field application of $7$ (blue), $8$ (red), and $9$ (black) mV/nm.}\label{visco}
\end{figure}

Figure \ref{visco} presents the shear ($\sigma$) and bulk ($\eta$) viscosity for $7$, $8$, and $9$ mV/nm applied field intensities. These values delineates the emergence of the main effects observed on phase diagram of Fig.\ref{diagram}. The field-dependence of viscosities is remarkable. However, it was not found a clear correlation to the phenomena reported elsewhere. In fact, a more careful analysis of the system dynamics need to be conducted to clarify this point. This important task is beyond the scope of the present work.

\section{Conclusions}

Despite the well known effects of iontophoresis, its mechanism of work is still poorly understood. Our results indicate that hydration effects related to vesicle formation could be tuned by constant external electrical fields. Our findings indicates that this is the underlying mechanism of iontophoresis. This technique usually employs constant or alternating ($5-10$ kHz) currents \cite{gollins2019retrospective} for therapeutic  purposes. However, we argue that DC field usage would also have booster diffusion effects due to vesicles creation and reincorporation. The application of static electric field on skin would have direct beneficial absorption effects on water-soluble topical agents as vitamin C\cite{VitaC2017topical}, vitamin D\cite{kechichian2018vitaminD}, and vitamin B12\cite{b12brescoll2015review}. Moreover, the dermal jet-injection of drugs by fine needles\cite{Dermojetsaray2005treatment,arbache2018activation} could also take advantage from the application of external electric field. 

It is the first report of large diversity of phenomena as function of field in the literature, to the best of our knowledge. The phase diagram including reversible/irreversible water-rich vesicle formation, phase transitions, and membrane disruption was presented and would be helpful to experimental verification of our findings. Interestingly,  electric field shielding effects are in the origin of observed phenomena following a general Arrhenius-like time dependence. At last, we report that field dependence of both shear and bulk viscosity appeared to play no role in this case.

\textbf{Acknowledgements.} The authors would like to thank the Brazilian agencies Conselho Nacional de Desenvolvimento Científico e Tecnológico (CNPq - 311146/2015-5) and Fundação de Amparo à Pesquisa do Estado de São Paulo (FAPESP - 2011/19924-2) for the financial support. The authors would also thank the computational resources provided by Centro Nacional de Processamento de Alto Desempenho em São Paulo (CENAPAD-UNICAMP) and Sistema de Computação Petaflópica (Tier 0) (Santos Dumont-LNCC) under  Sistema Nacional de Processamento de Alto Desempenho (SINAPAD) of the  Ministério da Ciência, Tecnologia e Inovação (MCTI).\\


\section{References}

\begin{thebibliography}{67}
\expandafter\ifx\csname natexlab\endcsname\relax\def\natexlab#1{#1}\fi
\providecommand{\url}[1]{\texttt{#1}}
\providecommand{\href}[2]{#2}
\providecommand{\path}[1]{#1}
\providecommand{\DOIprefix}{doi:}
\providecommand{\ArXivprefix}{arXiv:}
\providecommand{\URLprefix}{URL: }
\providecommand{\Pubmedprefix}{pmid:}
\providecommand{\doi}[1]{\href{http://dx.doi.org/#1}{\path{#1}}}
\providecommand{\Pubmed}[1]{\href{pmid:#1}{\path{#1}}}
\providecommand{\bibinfo}[2]{#2}
\ifx\xfnm\relax \def\xfnm[#1]{\unskip,\space#1}\fi
%Type = Inbook
\bibitem[{Hongbo and Maibach(2004)}]{hongbo}
\bibinfo{author}{Z.~Hongbo}, \bibinfo{author}{H.~I. Maibach},
  \bibinfo{title}{Dermatotoxicology}, \bibinfo{edition}{6th edition} ed.,
  \bibinfo{publisher}{CRC Press LCC}, \bibinfo{address}{USA},
  \bibinfo{year}{2004}.
%Type = Article
\bibitem[{Boer et~al.(2016)Boer, Duchnik, Maleszka, and
  Marchlewicz}]{boer2016structural}
\bibinfo{author}{M.~Boer}, \bibinfo{author}{E.~Duchnik},
  \bibinfo{author}{R.~Maleszka}, \bibinfo{author}{M.~Marchlewicz},
\newblock \bibinfo{title}{Structural and biophysical characteristics of human
  skin in maintaining proper epidermal barrier function},
\newblock \bibinfo{journal}{Adv. Dermatol. Allergol.} \bibinfo{volume}{33}
  (\bibinfo{year}{2016}) \bibinfo{pages}{1}.
%Type = Inbook
\bibitem[{Burkitt et~al.(2014)Burkitt, Young, and Heath}]{Burkitt2014}
\bibinfo{author}{H.~G. Burkitt}, \bibinfo{author}{B.~Young},
  \bibinfo{author}{J.~W. Heath}, \bibinfo{title}{Wheater's functional
  histology: a text and colour atlas}, \bibinfo{edition}{6th} ed.,
  \bibinfo{publisher}{Churchill Livingstone}, \bibinfo{year}{2014}, pp.
  \bibinfo{pages}{159--179}.
%Type = Book
\bibitem[{MD and MD(2013)}]{JamesG.MarksJr.2013}
\bibinfo{author}{J.~G. M.~J. MD}, \bibinfo{author}{J.~J.~M. MD},
  \bibinfo{title}{Lookingbill and Marks' Principles of Dermatology},
  \bibinfo{edition}{5e} ed., \bibinfo{publisher}{Saunders},
  \bibinfo{year}{2013}.
%Type = Article
\bibitem[{Harding et~al.(2013)Harding, Aho, and Bosko}]{Harding2013}
\bibinfo{author}{C.~R. Harding}, \bibinfo{author}{S.~Aho},
  \bibinfo{author}{C.~A. Bosko},
\newblock \bibinfo{title}{{Filaggrin - revisited}},
\newblock \bibinfo{journal}{Int. J. Cosmet. Sci.} \bibinfo{volume}{35}
  (\bibinfo{year}{2013}) \bibinfo{pages}{412--423}.
%Type = Book
\bibitem[{Elias(2016)}]{elias2016advances}
\bibinfo{author}{P.~M. Elias}, \bibinfo{title}{{Advances in Lipid Research:
  Skin Lipids}}, Advances in Lipid Research, \bibinfo{publisher}{Elsevier
  Science}, \bibinfo{year}{2016}. \URLprefix
  \url{https://books.google.com.br/books?id=rTeaBQAAQBAJ}.
%Type = Article
\bibitem[{Fox et~al.(2011)Fox, Gerber, Plessis, and
  Hamman}]{fox2011transdermal}
\bibinfo{author}{L.~T. Fox}, \bibinfo{author}{M.~Gerber},
  \bibinfo{author}{J.~D. Plessis}, \bibinfo{author}{J.~H. Hamman},
\newblock \bibinfo{title}{Transdermal drug delivery enhancement by compounds of
  natural origin},
\newblock \bibinfo{journal}{Molecules} \bibinfo{volume}{16}
  (\bibinfo{year}{2011}) \bibinfo{pages}{10507--10540}.
%Type = Article
\bibitem[{Herman and Herman(2015)}]{herman2015essential}
\bibinfo{author}{A.~Herman}, \bibinfo{author}{A.~P. Herman},
\newblock \bibinfo{title}{Essential oils and their constituents as skin
  penetration enhancer for transdermal drug delivery: a review},
\newblock \bibinfo{journal}{J. Pharm. Pharmacol.} \bibinfo{volume}{67}
  (\bibinfo{year}{2015}) \bibinfo{pages}{473--485}.
%Type = Article
\bibitem[{Dabrowska et~al.(2018)Dabrowska, Spano, Derler, Adlhart, Spencer, and
  Rossi}]{dkabrowska2018relationship}
\bibinfo{author}{A.~Dabrowska}, \bibinfo{author}{F.~Spano},
  \bibinfo{author}{S.~Derler}, \bibinfo{author}{C.~Adlhart},
  \bibinfo{author}{N.~Spencer}, \bibinfo{author}{R.~Rossi},
\newblock \bibinfo{title}{The relationship between skin function, barrier
  properties, and body-dependent factors},
\newblock \bibinfo{journal}{Skin Res Technol} \bibinfo{volume}{24}
  (\bibinfo{year}{2018}) \bibinfo{pages}{165--174}.
%Type = Article
\bibitem[{Pecoraro et~al.(2019)Pecoraro, Tutone, Hoffman, Hutter, Almerico, and
  Traynor}]{pecoraro2019predicting}
\bibinfo{author}{B.~Pecoraro}, \bibinfo{author}{M.~Tutone},
  \bibinfo{author}{E.~Hoffman}, \bibinfo{author}{V.~Hutter},
  \bibinfo{author}{A.~M. Almerico}, \bibinfo{author}{M.~Traynor},
\newblock \bibinfo{title}{Predicting skin permeability by means of
  computational approaches: Reliability and caveats in pharmaceutical studies},
\newblock \bibinfo{journal}{J. Chem. Inf. Model.}  (\bibinfo{year}{2019}).
%Type = Article
\bibitem[{Wosicka and Cal(2017)}]{Wosicka2017}
\bibinfo{author}{H.~Wosicka}, \bibinfo{author}{K.~Cal},
\newblock \bibinfo{title}{{Targeting to the hair follicles: Current status and
  potential}},
\newblock \bibinfo{journal}{J Dermatol Sci} \bibinfo{volume}{57}
  (\bibinfo{year}{2017}) \bibinfo{pages}{83--89}.
%Type = Article
\bibitem[{Schoellhammer et~al.(2014)Schoellhammer, Blankschtein, and
  Langer}]{schoellhammer2014skin}
\bibinfo{author}{C.~M. Schoellhammer}, \bibinfo{author}{D.~Blankschtein},
  \bibinfo{author}{R.~Langer},
\newblock \bibinfo{title}{Skin permeabilization for transdermal drug delivery:
  recent advances and future prospects},
\newblock \bibinfo{journal}{Expert Opin. Drug Deliv.} \bibinfo{volume}{11}
  (\bibinfo{year}{2014}) \bibinfo{pages}{393--407}.
%Type = Article
\bibitem[{Godin and Touitou(2007)}]{godin2007transdermal}
\bibinfo{author}{B.~Godin}, \bibinfo{author}{E.~Touitou},
\newblock \bibinfo{title}{Transdermal skin delivery: predictions for humans
  from in vivo, ex vivo and animal models},
\newblock \bibinfo{journal}{Adv. Drug Delivery Rev.} \bibinfo{volume}{59}
  (\bibinfo{year}{2007}) \bibinfo{pages}{1152--1161}.
%Type = Article
\bibitem[{{Batista Napotnik} and
  Miklav{\v{c}}i{\v{c}}(2018)}]{BatistaNapotnik2018}
\bibinfo{author}{T.~{Batista Napotnik}},
  \bibinfo{author}{D.~Miklav{\v{c}}i{\v{c}}},
\newblock \bibinfo{title}{{In vitro electroporation detection methods – An
  overview}},
\newblock \bibinfo{journal}{Bioelectrochemistry} \bibinfo{volume}{120}
  (\bibinfo{year}{2018}) \bibinfo{pages}{166--182}.
%Type = Article
\bibitem[{Kotnik et~al.(2015)Kotnik, Frey, Sack, Megli{\v{c}}, Peterka, and
  Miklav{\v{c}}i{\v{c}}}]{kotnik2015electroporation}
\bibinfo{author}{T.~Kotnik}, \bibinfo{author}{W.~Frey},
  \bibinfo{author}{M.~Sack}, \bibinfo{author}{S.~H. Megli{\v{c}}},
  \bibinfo{author}{M.~Peterka}, \bibinfo{author}{D.~Miklav{\v{c}}i{\v{c}}},
\newblock \bibinfo{title}{Electroporation-based applications in biotechnology},
\newblock \bibinfo{journal}{Trends Biotechnol.} \bibinfo{volume}{33}
  (\bibinfo{year}{2015}) \bibinfo{pages}{480--488}.
%Type = Article
\bibitem[{Rubinsky(2007)}]{rubinsky2007irreversible}
\bibinfo{author}{B.~Rubinsky},
\newblock \bibinfo{title}{Irreversible electroporation in medicine},
\newblock \bibinfo{journal}{Technol. Cancer Res. Treat.} \bibinfo{volume}{6}
  (\bibinfo{year}{2007}) \bibinfo{pages}{255--259}.
%Type = Article
\bibitem[{Potter and Heller(2018)}]{potter2018transfection}
\bibinfo{author}{H.~Potter}, \bibinfo{author}{R.~Heller},
\newblock \bibinfo{title}{Transfection by electroporation},
\newblock \bibinfo{journal}{Curr Protoc Mol Biol.} \bibinfo{volume}{121}
  (\bibinfo{year}{2018}) \bibinfo{pages}{9--3}.
%Type = Article
\bibitem[{Baba et~al.(2015)Baba, Takahara, and Mamitsuka}]{Baba2015}
\bibinfo{author}{H.~Baba}, \bibinfo{author}{J.-i. Takahara},
  \bibinfo{author}{H.~Mamitsuka},
\newblock \bibinfo{title}{In silico predictions of human skin permeability
  using nonlinear quantitative structure--property relationship models},
\newblock \bibinfo{journal}{Pharm. Res.} \bibinfo{volume}{32}
  (\bibinfo{year}{2015}) \bibinfo{pages}{2360--2371}.
%Type = Article
\bibitem[{Selzer et~al.(2015)Selzer, Neumann, Neumann, Kostka, Lehr, and
  Schaefer}]{Selzer2015}
\bibinfo{author}{D.~Selzer}, \bibinfo{author}{D.~Neumann},
  \bibinfo{author}{H.~Neumann}, \bibinfo{author}{K.-H. Kostka},
  \bibinfo{author}{C.-M. Lehr}, \bibinfo{author}{U.~F. Schaefer},
\newblock \bibinfo{title}{A strategy for in-silico prediction of skin
  absorption in man},
\newblock \bibinfo{journal}{Eur. J. Pharm. Biopharm.} \bibinfo{volume}{95}
  (\bibinfo{year}{2015}) \bibinfo{pages}{68--76}.
%Type = Article
\bibitem[{Hatanaka et~al.(2015)Hatanaka, Yoshida, Kadhum, Todo, and
  Sugibayashi}]{Hatanaka2015}
\bibinfo{author}{T.~Hatanaka}, \bibinfo{author}{S.~Yoshida},
  \bibinfo{author}{W.~R. Kadhum}, \bibinfo{author}{H.~Todo},
  \bibinfo{author}{K.~Sugibayashi},
\newblock \bibinfo{title}{In silico estimation of skin concentration following
  the dermal exposure to chemicals},
\newblock \bibinfo{journal}{Pharm. Res.} \bibinfo{volume}{32}
  (\bibinfo{year}{2015}) \bibinfo{pages}{3965--3974}.
%Type = Article
\bibitem[{Verheyen et~al.(2017)Verheyen, Braeken, Van~Deun, and
  Van~Miert}]{Verheyen2017}
\bibinfo{author}{G.~Verheyen}, \bibinfo{author}{E.~Braeken},
  \bibinfo{author}{K.~Van~Deun}, \bibinfo{author}{S.~Van~Miert},
\newblock \bibinfo{title}{Evaluation of in silico tools to predict the skin
  sensitization potential of chemicals},
\newblock \bibinfo{journal}{SAR QSAR Environ. Res.} \bibinfo{volume}{28}
  (\bibinfo{year}{2017}) \bibinfo{pages}{59--73}.
%Type = Article
\bibitem[{Marrink et~al.(2019)Marrink, Corradi, Souza, Ing\'olfsson, Tieleman,
  and Sansom}]{marrink2019computational}
\bibinfo{author}{S.~J. Marrink}, \bibinfo{author}{V.~Corradi},
  \bibinfo{author}{P.~C. Souza}, \bibinfo{author}{H.~I. Ing\'olfsson},
  \bibinfo{author}{D.~P. Tieleman}, \bibinfo{author}{M.~S. Sansom},
\newblock \bibinfo{title}{Computational modeling of realistic cell membranes},
\newblock \bibinfo{journal}{Chem. Rev.}  (\bibinfo{year}{2019}).
%Type = Article
\bibitem[{Machado et~al.(2016)Machado, {Dos Santos}, Carvalho, Singh, {Tellez
  Soto}, Azoia, Cavaco-Paulo, Martin, and Favero}]{Machado2016}
\bibinfo{author}{N.~C.~F. Machado}, \bibinfo{author}{L.~{Dos Santos}},
  \bibinfo{author}{B.~G. Carvalho}, \bibinfo{author}{P.~Singh},
  \bibinfo{author}{C.~A. {Tellez Soto}}, \bibinfo{author}{N.~G. Azoia},
  \bibinfo{author}{A.~Cavaco-Paulo}, \bibinfo{author}{A.~A. Martin},
  \bibinfo{author}{P.~P. Favero},
\newblock \bibinfo{title}{{Assessment of penetration of Ascorbyl
  Tetraisopalmitate into biological membranes by molecular dynamics.}},
\newblock \bibinfo{journal}{Comput. Biol. Med.} \bibinfo{volume}{75}
  (\bibinfo{year}{2016}) \bibinfo{pages}{151--159}.
%Type = Article
\bibitem[{Rocco et~al.(2017)Rocco, Cilurzo, Minghetti, Vistoli, and
  Pedretti}]{rocco2017molecular}
\bibinfo{author}{P.~Rocco}, \bibinfo{author}{F.~Cilurzo},
  \bibinfo{author}{P.~Minghetti}, \bibinfo{author}{G.~Vistoli},
  \bibinfo{author}{A.~Pedretti},
\newblock \bibinfo{title}{Molecular dynamics as a tool for in silico screening
  of skin permeability},
\newblock \bibinfo{journal}{Eur. J. Pharm. Sci.} \bibinfo{volume}{106}
  (\bibinfo{year}{2017}) \bibinfo{pages}{328--335}.
%Type = Article
\bibitem[{Gupta and Rai(2018)}]{gupta2018electroporation}
\bibinfo{author}{R.~Gupta}, \bibinfo{author}{B.~Rai},
\newblock \bibinfo{title}{Electroporation of skin stratum corneum lipid bilayer
  and molecular mechanism of drug transport: A molecular dynamics study},
\newblock \bibinfo{journal}{Langmuir} \bibinfo{volume}{34}
  (\bibinfo{year}{2018}) \bibinfo{pages}{5860--5870}.
%Type = Article
\bibitem[{Groves et~al.(2000)Groves, Boxer, and McConnell}]{Groves2000}
\bibinfo{author}{J.~T. Groves}, \bibinfo{author}{S.~G. Boxer},
  \bibinfo{author}{H.~M. McConnell},
\newblock \bibinfo{title}{{Lateral Reorganization of Fluid Lipid Membranes in
  Response to the Electric Field Produced by a Buried Charge}},
\newblock \bibinfo{journal}{J. Phys. Chem. B} \bibinfo{volume}{104}
  (\bibinfo{year}{2000}) \bibinfo{pages}{11409--11415}.
%Type = Incollection
\bibitem[{Tarek(2017)}]{tarek2017atomistic}
\bibinfo{author}{M.~Tarek},
\newblock \bibinfo{title}{Atomistic simulations of electroporation of model
  cell membranes},
\newblock in: \bibinfo{booktitle}{Transport Across Natural and Modified
  Biological Membranes and its Implications in Physiology and Therapy},
  \bibinfo{publisher}{Springer}, \bibinfo{year}{2017}, pp.
  \bibinfo{pages}{1--15}.
%Type = Article
\bibitem[{Polak et~al.(2014)Polak, Tarek, Tom{\v{s}}i{\v{c}}, Valant, Ulrih,
  Jamnik, Kramar, and Miklav{\v{c}}i{\v{c}}}]{Polak2014}
\bibinfo{author}{A.~Polak}, \bibinfo{author}{M.~Tarek},
  \bibinfo{author}{M.~Tom{\v{s}}i{\v{c}}}, \bibinfo{author}{J.~Valant},
  \bibinfo{author}{N.~P. Ulrih}, \bibinfo{author}{A.~Jamnik},
  \bibinfo{author}{P.~Kramar}, \bibinfo{author}{D.~Miklav{\v{c}}i{\v{c}}},
\newblock \bibinfo{title}{{Electroporation of archaeal lipid membranes using MD
  simulations}},
\newblock \bibinfo{journal}{Bioelectrochemistry} \bibinfo{volume}{100}
  (\bibinfo{year}{2014}) \bibinfo{pages}{18--26}.
%Type = Article
\bibitem[{Koronkiewicz and Kalinowski(2004)}]{Koronkiewicz2004}
\bibinfo{author}{S.~Koronkiewicz}, \bibinfo{author}{S.~Kalinowski},
\newblock \bibinfo{title}{{Influence of cholesterol on electroporation of
  bilayer lipid membranes: chronopotentiometric studies}},
\newblock \bibinfo{journal}{Biochim. Biophys. Acta, Biomembr.}
  \bibinfo{volume}{1661} (\bibinfo{year}{2004}) \bibinfo{pages}{196--203}.
%Type = Article
\bibitem[{Casciola et~al.(2014)Casciola, Bonhenry, Liberti, Apollonio, and
  Tarek}]{Casciola2014}
\bibinfo{author}{M.~Casciola}, \bibinfo{author}{D.~Bonhenry},
  \bibinfo{author}{M.~Liberti}, \bibinfo{author}{F.~Apollonio},
  \bibinfo{author}{M.~Tarek},
\newblock \bibinfo{title}{{A molecular dynamic study of cholesterol rich lipid
  membranes: comparison of electroporation protocols}},
\newblock \bibinfo{journal}{Bioelectrochemistry} \bibinfo{volume}{100}
  (\bibinfo{year}{2014}) \bibinfo{pages}{11--17}.
%Type = Article
\bibitem[{Bennett et~al.(2009)Bennett, MacCallum, and Tieleman}]{Bennett2009}
\bibinfo{author}{W.~F.~D. Bennett}, \bibinfo{author}{J.~L. MacCallum},
  \bibinfo{author}{D.~P. Tieleman},
\newblock \bibinfo{title}{{Thermodynamic Analysis of the Effect of Cholesterol
  on Dipalmitoylphosphatidylcholine Lipid Membranes}},
\newblock \bibinfo{journal}{JACS} \bibinfo{volume}{131} (\bibinfo{year}{2009})
  \bibinfo{pages}{1972--1978}.
%Type = Article
\bibitem[{Fern{\'{a}}ndez et~al.(2010)Fern{\'{a}}ndez, Marshall, Sagu{\'{e}}s,
  and Reigada}]{Fernandez2010}
\bibinfo{author}{M.~L. Fern{\'{a}}ndez}, \bibinfo{author}{G.~Marshall},
  \bibinfo{author}{F.~Sagu{\'{e}}s}, \bibinfo{author}{R.~Reigada},
\newblock \bibinfo{title}{{Structural and Kinetic Molecular Dynamics Study of
  Electroporation in Cholesterol-Containing Bilayers}},
\newblock \bibinfo{journal}{J. Phys. Chem. B} \bibinfo{volume}{114}
  (\bibinfo{year}{2010}) \bibinfo{pages}{6855--6865}.
%Type = Incollection
\bibitem[{Singhal and Kalia(2017)}]{singhal2017iontophoresis}
\bibinfo{author}{M.~Singhal}, \bibinfo{author}{Y.~N. Kalia},
\newblock \bibinfo{title}{Iontophoresis and electroporation},
\newblock in: \bibinfo{booktitle}{Skin Permeation and Disposition of
  Therapeutic and Cosmeceutical Compounds}, \bibinfo{publisher}{Springer},
  \bibinfo{year}{2017}, pp. \bibinfo{pages}{165--182}.
%Type = Article
\bibitem[{Prausnitz and Langer(2008)}]{prausnitz2008transdermal}
\bibinfo{author}{M.~R. Prausnitz}, \bibinfo{author}{R.~Langer},
\newblock \bibinfo{title}{Transdermal drug delivery},
\newblock \bibinfo{journal}{Nat. Biotechnol.} \bibinfo{volume}{26}
  (\bibinfo{year}{2008}) \bibinfo{pages}{1261}.
%Type = Article
\bibitem[{Gollins et~al.(2019)Gollins, Carpenter, Steen, Bulinski, and
  Mahendran}]{gollins2019retrospective}
\bibinfo{author}{C.~Gollins}, \bibinfo{author}{A.~Carpenter},
  \bibinfo{author}{C.~Steen}, \bibinfo{author}{H.~Bulinski},
  \bibinfo{author}{R.~Mahendran},
\newblock \bibinfo{title}{A retrospective analysis of the use of tap water
  iontophoresis for focal hyperhidrosis at a district general hospital: The
  patients’ perspective},
\newblock \bibinfo{journal}{J Dermatolog Treat.}  (\bibinfo{year}{2019})
  \bibinfo{pages}{1--9}.
%Type = Incollection
\bibitem[{Wallis(2019)}]{wallis2019diagnosis}
\bibinfo{author}{C.~Wallis},
\newblock \bibinfo{title}{Diagnosis and presentation of cystic fibrosis},
\newblock in: \bibinfo{booktitle}{Kendig's Disorders of the Respiratory Tract
  in Children (Ninth Edition)}, \bibinfo{publisher}{Elsevier},
  \bibinfo{year}{2019}, pp. \bibinfo{pages}{769--776}.
%Type = Article
\bibitem[{Liu and Liu(2017)}]{liu2017iontophoretic}
\bibinfo{author}{F.~Liu}, \bibinfo{author}{L.~Liu},
\newblock \bibinfo{title}{Iontophoretic delivery of lignocaine in healthy
  volunteers: effect of concentration of epinephrine on local anesthesia},
\newblock \bibinfo{journal}{Biomed Res India} \bibinfo{volume}{28}
  (\bibinfo{year}{2017}) \bibinfo{pages}{185--191}.
%Type = Article
\bibitem[{Byrne et~al.(2018)Byrne, Yeh, and DeSimone}]{byrne2018use}
\bibinfo{author}{J.~D. Byrne}, \bibinfo{author}{J.~J. Yeh},
  \bibinfo{author}{J.~M. DeSimone},
\newblock \bibinfo{title}{Use of iontophoresis for the treatment of cancer},
\newblock \bibinfo{journal}{J. Controlled Release}  (\bibinfo{year}{2018}).
%Type = Incollection
\bibitem[{Gratieri and Kalia(2017)}]{gratieri2017iontophoresis}
\bibinfo{author}{T.~Gratieri}, \bibinfo{author}{Y.~N. Kalia},
\newblock \bibinfo{title}{Iontophoresis: Basic principles},
\newblock in: \bibinfo{booktitle}{Percutaneous Penetration Enhancers Physical
  Methods in Penetration Enhancement}, \bibinfo{publisher}{Springer},
  \bibinfo{year}{2017}, pp. \bibinfo{pages}{61--65}.
%Type = Article
\bibitem[{Kassan et~al.(1996)Kassan, Lynch, and Stiller}]{kassan1996physical}
\bibinfo{author}{D.~G. Kassan}, \bibinfo{author}{A.~M. Lynch},
  \bibinfo{author}{M.~J. Stiller},
\newblock \bibinfo{title}{Physical enhancement of dermatologic drug delivery:
  iontophoresis and phonophoresis},
\newblock \bibinfo{journal}{J Am Acad Dermatol} \bibinfo{volume}{34}
  (\bibinfo{year}{1996}) \bibinfo{pages}{657--666}.
%Type = Article
\bibitem[{Banga et~al.(1999)Banga, Bose, and Ghosh}]{banga1999iontophoresis}
\bibinfo{author}{A.~K. Banga}, \bibinfo{author}{S.~Bose},
  \bibinfo{author}{T.~K. Ghosh},
\newblock \bibinfo{title}{Iontophoresis and electroporation: comparisons and
  contrasts},
\newblock \bibinfo{journal}{Int. J. Pharm.} \bibinfo{volume}{179}
  (\bibinfo{year}{1999}) \bibinfo{pages}{1--19}.
%Type = Article
\bibitem[{Jadoul et~al.(1999)Jadoul, Bouwstra, and Preat}]{jadoul1999effects}
\bibinfo{author}{A.~Jadoul}, \bibinfo{author}{J.~Bouwstra},
  \bibinfo{author}{V.~Preat},
\newblock \bibinfo{title}{Effects of iontophoresis and electroporation on the
  stratum corneum: review of the biophysical studies},
\newblock \bibinfo{journal}{Adv. Drug Delivery Rev.} \bibinfo{volume}{35}
  (\bibinfo{year}{1999}) \bibinfo{pages}{89--105}.
%Type = Article
\bibitem[{Stewart et~al.(2018)Stewart, Langer, and
  Jensen}]{stewart2018intracellular}
\bibinfo{author}{M.~P. Stewart}, \bibinfo{author}{R.~Langer},
  \bibinfo{author}{K.~F. Jensen},
\newblock \bibinfo{title}{Intracellular delivery by membrane disruption:
  mechanisms, strategies, and concepts},
\newblock \bibinfo{journal}{Chemical reviews} \bibinfo{volume}{118}
  (\bibinfo{year}{2018}) \bibinfo{pages}{7409--7531}.
%Type = Article
\bibitem[{Marrink et~al.(2004)Marrink, de~Vries, and Mark}]{Marrink2004}
\bibinfo{author}{S.~J. Marrink}, \bibinfo{author}{A.~H. de~Vries},
  \bibinfo{author}{A.~E. Mark},
\newblock \bibinfo{title}{{Coarse Grained Model for Semiquantitative Lipid
  Simulations}},
\newblock \bibinfo{journal}{J. Phys. Chem. B} \bibinfo{volume}{108}
  (\bibinfo{year}{2004}) \bibinfo{pages}{750--760}.
%Type = Article
\bibitem[{Marrink et~al.(2007)Marrink, Risselada, Yefimov, Tieleman, and
  de~Vries}]{Marrink2007}
\bibinfo{author}{S.~J. Marrink}, \bibinfo{author}{H.~J. Risselada},
  \bibinfo{author}{S.~Yefimov}, \bibinfo{author}{D.~P. Tieleman},
  \bibinfo{author}{A.~H. de~Vries},
\newblock \bibinfo{title}{{The MARTINI Force Field: Coarse Grained Model for
  Biomolecular Simulations}},
\newblock \bibinfo{journal}{J. Phys. Chem. B} \bibinfo{volume}{111}
  (\bibinfo{year}{2007}) \bibinfo{pages}{7812--7824}.
%Type = Article
\bibitem[{Ingolfsson et~al.(2014)Ingolfsson, Lopez, Uusitalo, de~Jong, Gopal,
  Periole, and Marrink}]{Ingolfsson2014a}
\bibinfo{author}{H.~I. Ingolfsson}, \bibinfo{author}{C.~A. Lopez},
  \bibinfo{author}{J.~J. Uusitalo}, \bibinfo{author}{D.~H. de~Jong},
  \bibinfo{author}{S.~M. Gopal}, \bibinfo{author}{X.~Periole},
  \bibinfo{author}{S.~J. Marrink},
\newblock \bibinfo{title}{{The power of coarse graining in biomolecular
  simulations.}},
\newblock \bibinfo{journal}{Wiley Interdiscip. Rev.: Comput. Mol. Sci.}
  \bibinfo{volume}{4} (\bibinfo{year}{2014}) \bibinfo{pages}{225--248}.
%Type = Article
\bibitem[{{Van Der Spoel} et~al.(2005){Van Der Spoel}, Lindahl, Hess, Groenhof,
  Mark, and Berendsen}]{VanDerSpoel2005}
\bibinfo{author}{D.~{Van Der Spoel}}, \bibinfo{author}{E.~Lindahl},
  \bibinfo{author}{B.~Hess}, \bibinfo{author}{G.~Groenhof},
  \bibinfo{author}{A.~E. Mark}, \bibinfo{author}{H.~J.~C. Berendsen},
\newblock \bibinfo{title}{{GROMACS: Fast, flexible, and free}},
\newblock \bibinfo{journal}{J. Comput. Chem.} \bibinfo{volume}{26}
  (\bibinfo{year}{2005}) \bibinfo{pages}{1701--1718}.
%Type = Article
\bibitem[{Abraham et~al.(2015)Abraham, Murtola, Schulz, P{\'{a}}ll, Smith,
  Hess, and Lindahl}]{Abraham2015}
\bibinfo{author}{M.~J. Abraham}, \bibinfo{author}{T.~Murtola},
  \bibinfo{author}{R.~Schulz}, \bibinfo{author}{S.~P{\'{a}}ll},
  \bibinfo{author}{J.~C. Smith}, \bibinfo{author}{B.~Hess},
  \bibinfo{author}{E.~Lindahl},
\newblock \bibinfo{title}{{GROMACS: High performance molecular simulations
  through multi-level parallelism from laptops to supercomputers}},
\newblock \bibinfo{journal}{SoftwareX} \bibinfo{volume}{1-2}
  (\bibinfo{year}{2015}) \bibinfo{pages}{19--25}.
%Type = Article
\bibitem[{Berendsen et~al.(1995)Berendsen, van~der Spoel, and van
  Drunen}]{Berendsen1995}
\bibinfo{author}{H.~Berendsen}, \bibinfo{author}{D.~van~der Spoel},
  \bibinfo{author}{R.~van Drunen},
\newblock \bibinfo{title}{{GROMACS: A message-passing parallel molecular
  dynamics implementation}},
\newblock \bibinfo{journal}{Comput. Phys. Commun.} \bibinfo{volume}{91}
  (\bibinfo{year}{1995}) \bibinfo{pages}{43--56}.
%Type = Article
\bibitem[{Yesylevskyy et~al.(2010)Yesylevskyy, Schäfer, Sengupta, and
  Marrink}]{Yesylevskyy2010}
\bibinfo{author}{S.~O. Yesylevskyy}, \bibinfo{author}{L.~V. Schäfer},
  \bibinfo{author}{D.~Sengupta}, \bibinfo{author}{S.~J. Marrink},
\newblock \bibinfo{title}{Polarizable water model for the coarse-grained
  martini force field},
\newblock \bibinfo{journal}{PLOS Computational Biology} \bibinfo{volume}{6}
  (\bibinfo{year}{2010}) \bibinfo{pages}{1--17}.
%Type = Article
\bibitem[{Wassenaar et~al.(2015)Wassenaar, Ing{\'{o}}lfsson, B{\"{o}}ckmann,
  Tieleman, and Marrink}]{Wassenaar2015}
\bibinfo{author}{T.~A. Wassenaar}, \bibinfo{author}{H.~I. Ing{\'{o}}lfsson},
  \bibinfo{author}{R.~A. B{\"{o}}ckmann}, \bibinfo{author}{D.~P. Tieleman},
  \bibinfo{author}{S.~J. Marrink},
\newblock \bibinfo{title}{{Computational Lipidomics with insane : A Versatile
  Tool for Generating Custom Membranes for Molecular Simulations}},
\newblock \bibinfo{journal}{J. Chem. Theory Comput.} \bibinfo{volume}{11}
  (\bibinfo{year}{2015}) \bibinfo{pages}{2144--2155}.
%Type = Article
\bibitem[{Hockney et~al.(1974)Hockney, Goel, and Eastwood}]{Hockney1974}
\bibinfo{author}{R.~Hockney}, \bibinfo{author}{S.~Goel},
  \bibinfo{author}{J.~Eastwood},
\newblock \bibinfo{title}{{Quiet high-resolution computer models of a plasma}},
\newblock \bibinfo{journal}{J. Comput. Phys.} \bibinfo{volume}{14}
  (\bibinfo{year}{1974}) \bibinfo{pages}{148--158}.
%Type = Article
\bibitem[{Verlet(1967)}]{Verlet1967}
\bibinfo{author}{L.~Verlet},
\newblock \bibinfo{title}{{Computer "Experiments" on Classical Fluids. I.
  Thermodynamical Properties of Lennard-Jones Molecules}},
\newblock \bibinfo{journal}{Phys. Rev.} \bibinfo{volume}{159}
  (\bibinfo{year}{1967}) \bibinfo{pages}{98--103}.
%Type = Article
\bibitem[{Dilger et~al.(1979)Dilger, McLaughlin, McIntosh, and
  Simon}]{Dilger1196}
\bibinfo{author}{J.~Dilger}, \bibinfo{author}{S.~McLaughlin},
  \bibinfo{author}{T.~McIntosh}, \bibinfo{author}{S.~Simon},
\newblock \bibinfo{title}{The dielectric constant of phospholipid bilayers and
  the permeability of membranes to ions},
\newblock \bibinfo{journal}{Science} \bibinfo{volume}{206}
  (\bibinfo{year}{1979}) \bibinfo{pages}{1196--1198}.
%Type = Article
\bibitem[{Gramse et~al.(2013)Gramse, Dols-Perez, Edwards, Fumagalli, and
  Gomila}]{Gramse2013}
\bibinfo{author}{G.~Gramse}, \bibinfo{author}{A.~Dols-Perez},
  \bibinfo{author}{M.~A. Edwards}, \bibinfo{author}{L.~Fumagalli},
  \bibinfo{author}{G.~Gomila},
\newblock \bibinfo{title}{{Nanoscale measurement of the dielectric constant of
  supported lipid bilayers in aqueous solutions with electrostatic force
  microscopy}},
\newblock \bibinfo{journal}{Biophys. J.} \bibinfo{volume}{104}
  (\bibinfo{year}{2013}) \bibinfo{pages}{1257--1262}.
%Type = Article
\bibitem[{Berendsen et~al.(1984)Berendsen, Postma, van Gunsteren, DiNola, and
  Haak}]{Berendsen1984}
\bibinfo{author}{H.~J.~C. Berendsen}, \bibinfo{author}{J.~P.~M. Postma},
  \bibinfo{author}{W.~F. van Gunsteren}, \bibinfo{author}{A.~DiNola},
  \bibinfo{author}{J.~R. Haak},
\newblock \bibinfo{title}{{Molecular dynamics with coupling to an external
  bath}},
\newblock \bibinfo{journal}{J. Chem. Phys.} \bibinfo{volume}{81}
  (\bibinfo{year}{1984}) \bibinfo{pages}{3684--3690}.
%Type = Article
\bibitem[{Bussi et~al.(2007)Bussi, Donadio, and Parrinello}]{Bussi2007}
\bibinfo{author}{G.~Bussi}, \bibinfo{author}{D.~Donadio},
  \bibinfo{author}{M.~Parrinello},
\newblock \bibinfo{title}{{Canonical sampling through velocity rescaling.}},
\newblock \bibinfo{journal}{J Chem Phys.} \bibinfo{volume}{126}
  (\bibinfo{year}{2007}) \bibinfo{pages}{014101}.
%Type = Article
\bibitem[{Parrinello and Rahman(1981)}]{parrinello1981polymorphic}
\bibinfo{author}{M.~Parrinello}, \bibinfo{author}{A.~Rahman},
\newblock \bibinfo{title}{Polymorphic transitions in single crystals: A new
  molecular dynamics method},
\newblock \bibinfo{journal}{J. Appl. Phys.} \bibinfo{volume}{52}
  (\bibinfo{year}{1981}) \bibinfo{pages}{7182--7190}.
%Type = Article
\bibitem[{Caleman and Van Der~Spoel(2008)}]{caleman2008picosecond}
\bibinfo{author}{C.~Caleman}, \bibinfo{author}{D.~Van Der~Spoel},
\newblock \bibinfo{title}{Picosecond melting of ice by an infrared laser pulse:
  A simulation study},
\newblock \bibinfo{journal}{Angew. Chem. Int. Ed.} \bibinfo{volume}{47}
  (\bibinfo{year}{2008}) \bibinfo{pages}{1417--1420}.
%Type = Article
\bibitem[{Humphrey et~al.(1996)Humphrey, Dalke, and Schulten}]{humphrey1996vmd}
\bibinfo{author}{W.~Humphrey}, \bibinfo{author}{A.~Dalke},
  \bibinfo{author}{K.~Schulten},
\newblock \bibinfo{title}{Vmd: visual molecular dynamics},
\newblock \bibinfo{journal}{J. Mol. Graphics} \bibinfo{volume}{14}
  (\bibinfo{year}{1996}) \bibinfo{pages}{33--38}.
%Type = Book
\bibitem[{Heald and Marion(1995)}]{heald1995classical}
\bibinfo{author}{M.~A. Heald}, \bibinfo{author}{J.~B. Marion},
  \bibinfo{title}{Classical electromagnetic radiation},
  \bibinfo{publisher}{Saunders College Pub.}, \bibinfo{year}{1995}.
  \bibinfo{note}{Using relative dielectric constant value of 14 from our
  model}.
%Type = Article
\bibitem[{Pipe and McKinley(2009)}]{pipe2009microfluidic}
\bibinfo{author}{C.~J. Pipe}, \bibinfo{author}{G.~H. McKinley},
\newblock \bibinfo{title}{Microfluidic rheometry},
\newblock \bibinfo{journal}{Mech. Res. Commun.} \bibinfo{volume}{36}
  (\bibinfo{year}{2009}) \bibinfo{pages}{110--120}.
%Type = Article
\bibitem[{Al-Niaimi and Chiang(2017)}]{VitaC2017topical}
\bibinfo{author}{F.~Al-Niaimi}, \bibinfo{author}{N.~Y.~Z. Chiang},
\newblock \bibinfo{title}{Topical vitamin c and the skin: mechanisms of action
  and clinical applications},
\newblock \bibinfo{journal}{J Clin Aesthet Dermatol} \bibinfo{volume}{10}
  (\bibinfo{year}{2017}) \bibinfo{pages}{14}.
%Type = Article
\bibitem[{Kechichian and Ezzedine(2018)}]{kechichian2018vitaminD}
\bibinfo{author}{E.~Kechichian}, \bibinfo{author}{K.~Ezzedine},
\newblock \bibinfo{title}{Vitamin d and the skin: an update for
  dermatologists},
\newblock \bibinfo{journal}{Am J Clin Dermatol} \bibinfo{volume}{19}
  (\bibinfo{year}{2018}) \bibinfo{pages}{223--235}.
%Type = Article
\bibitem[{Brescoll and Daveluy(2015)}]{b12brescoll2015review}
\bibinfo{author}{J.~Brescoll}, \bibinfo{author}{S.~Daveluy},
\newblock \bibinfo{title}{A review of vitamin b12 in dermatology},
\newblock \bibinfo{journal}{Am J Clin Dermatol} \bibinfo{volume}{16}
  (\bibinfo{year}{2015}) \bibinfo{pages}{27--33}.
%Type = Article
\bibitem[{Saray and G{\"u}le{\c{c}}(2005)}]{Dermojetsaray2005treatment}
\bibinfo{author}{Y.~Saray}, \bibinfo{author}{A.~T. G{\"u}le{\c{c}}},
\newblock \bibinfo{title}{Treatment of keloids and hypertrophic scars with
  dermojet injections of bleomycin: a preliminary study},
\newblock \bibinfo{journal}{Int J Dermatol} \bibinfo{volume}{44}
  (\bibinfo{year}{2005}) \bibinfo{pages}{777--784}.
%Type = Article
\bibitem[{Arbache et~al.(2018)Arbache, Roth, Steiner, Breunig, Michalany,
  Arbache, de~Souza, and Hirata}]{arbache2018activation}
\bibinfo{author}{S.~Arbache}, \bibinfo{author}{D.~Roth},
  \bibinfo{author}{D.~Steiner}, \bibinfo{author}{J.~Breunig},
  \bibinfo{author}{N.~S. Michalany}, \bibinfo{author}{S.~T. Arbache},
  \bibinfo{author}{L.~G. de~Souza}, \bibinfo{author}{S.~H. Hirata},
\newblock \bibinfo{title}{Activation of melanocytes in idiopathic guttate
  hypomelanosis after 5-fluorouracil infusion using a tattoo machine:
  Preliminary analysis of a randomized, split-body, single blinded, placebo
  controlled clinical trial},
\newblock \bibinfo{journal}{J Am Acad Dermatol} \bibinfo{volume}{78}
  (\bibinfo{year}{2018}) \bibinfo{pages}{212--215}.

\end{thebibliography}
\end{document}